\documentclass[12pt]{iopart}
\usepackage{geometry}                
\usepackage{graphicx}
\usepackage{bm}
\usepackage{amssymb}
\usepackage{mathrsfs} 
\usepackage{setstack}
\usepackage{epstopdf}
\def\TODAY{16 April 2010}
\begin{document}
\title{General polarization modes for the Rosen gravitational wave}
\author{Bethan Cropp}
\address{School of Mathematics, Statistics, and Operations Research, \\
Victoria University of Wellington, \\
Wellington, New Zealand
}
\author{Matt Visser}
\address{School of Mathematics, Statistics, and Operations Research, \\
Victoria University of Wellington, \\
Wellington, New Zealand
}

\date{\TODAY;  \LaTeX-ed \today}                                           
\begin{abstract}
Strong-field gravitational plane waves are often represented in either the Rosen or Brinkmann forms.  While these two metric ansatz\"e are related by a coordinate transformation, so that they should describe essentially the same physics, they rather puzzlingly seem to treat polarization states quite differently. Both ansatz\"e deal equally well with $+$ and $\times$ linear polarizations, but there is a qualitative difference in the way they deal with circular, elliptic,  and  more general polarization states. In this article we will develop a general formalism for dealing with arbitrary polarization states in the Rosen form of the gravitational wave metric, representing an arbitrary polarization by a trajectory in a suitably defined two dimensional hyperbolic plane. 
$+$~polarization; $\times$~polarization; circular polarization; general polarization.
\\
\\

\end{abstract}
\maketitle
\clearpage
\def\d{{\mathrm{d}}}
\newcommand{\scri}{\mathscr{I}}
\newcommand{\sun}{\ensuremath{\odot}}
\def\J{{\mathscr{J}}}
\def\sech{{\mathrm{sech}}}
\def\T{{\mathcal{T}}}
\def\tr{{\mathrm{tr}}}


\section{Introduction}

Strong field gravitational plane waves (a restriction of the more general $pp$~waves) are commonly represented in either the Brinkmann form~\cite{Brinkmann} 
\begin{equation}
\d s^2 = - 2 \,\d u \,\d v + \left\{ H_{AB}(u)\, x^A\, x^B\right\} \, \d u^2 + \d x^2 + \d y^2,
\end{equation}
or the Rosen form~\cite{Rosen}
\begin{equation}
\d s^2 = - 2 \,\d u \,\d v + g_{AB}(u)\;\d x^A\,\d x^B,
\end{equation}
where $x^A = \{ x, y \}$. While these two metric ansatz\"e are related by a coordinate transformation~\cite{D'Inverno, exact}, and so must describe essentially the same physics, there is a qualitative difference in how they treat circular polarization, elliptic polarization, and more general states of polarization. 
We shall investigate this puzzle in detail, and will ultimately demonstrate a clean way of putting a general polarization state into the Rosen form.

\section{Brinkmann form}

Consider the general $pp$ spacetime geometry~\cite{exact, Penrose1, Penrose2, pp} 
\begin{equation}
\d s^2 = - 2\,  \d u \, \d v + H(u,x,y) \, \d u^2+ \d x^2 + \d y^2. 
\end{equation}
It is then a standard result that the only nonzero component of the  Ricci tensor is
\begin{equation}
R_{uu}= -{1\over2} \left\{ \partial_x^2 H(u,x,y) + \partial_y^2H(u,x,y) \right\}.
\end{equation}
Furthermore  (up to the usual index symmetries) the only non-zero components of the Riemann tensor are of the form $R_{uAuB}$. Specifically:
\begin{equation}
R_{uxux} = -{1\over2} \partial_x^2 H(u,x,y); 
\end{equation}
\begin{equation}
R_{uxuy} = -{1\over2} \partial_x \partial_y H(u,x,y) ;
\end{equation}
\begin{equation}
R_{uyuy} = -{1\over2} \partial_y^2 H(u,x,y) . 
\end{equation}
Polarization modes are characterized by the relative motion of nearby timelike geodesics under the influence of the imposed $pp$-wave spacetime; as such the polarization modes are sensitive to the individual nonzero components of the Riemann tensor: $R_{uxux}$, $R_{uxuy}$ and $ R_{uyuy} $.  This is the basis of the discussion in \S 17.2 of Griffiths and Podolsk\'y~\cite{Griffiths},  which has the net effect that one can safely carry over one's intuition from weak field gravitational waves to strong field gravitational waves in the Brinkmann form.

Gravitational plane waves, as opposed to the more general $pp$-waves~\cite{Penrose1, Penrose2, pp}, can be characterized by the fact that $H(u,x,y)$ is a quadratic function of the coordinates $x$ and $y$~\cite{pp}.  (See also~\cite{Brinkmann,  D'Inverno}.) Then
\begin{equation}
\fl \d s^2 = - 2 \,\d u \,\d v + \left\{ H_{AB}(u)\, x^A\, x^B\right\} \, \d u^2 + \d x^2 + \d y^2.
\end{equation}
The vacuum gravitational plane wave for arbitrary polarization can now be written down by inspection:
\begin{equation}
\label{E:1}
\fl \d s^2 = - 2\,  \d u \, \d v + \left\{ [ x^2-y^2] \, H_+(u) + 2xy \, H_\times(u) \right\} \, \d u^2+ \d x^2 + \d y^2. 
\end{equation}
Equivalently
\begin{equation}
\label{E:2}
\fl \d s^2 = - 2\,  \d u \, \d v +  r^2\left\{  \cos(2\phi) \, H_+(u) + \sin(2\phi) \, H_\times(u) \right\} \, \d u^2+ \d r^2 + r^2\,\d \phi^2. 
\end{equation}
In either of these two forms for the metric the two polarization modes are explicitly seen to decouple and to superimpose linearly --- quite similarly to the situation in Maxwell electromagnetism --- by choosing $H_+(u)$ and $H_\times(u)$ appropriately we can construct not just $+$ and $\times$ polarized waves, but also circular polarization, elliptic polarization, and even more general polarization states. \emph{It is this decoupling that fails in the Rosen form of the metric, and which ultimately is the source of the puzzle}.

\section{Rosen form}

\subsection{Most general Rosen form}

The ``most general''  form of the Rosen metric is~\cite{Rosen, exact, pp}
\begin{equation}
\d s^2 = - 2 \, \d u\, \d v + g_{AB}(u)\;\d x^A\,\d x^B,
\end{equation}
where $x^A = \{ x,y\}$. 
It is a standard result~\cite{exact}, quite easily checked using symbolic manipulation systems such as {\sf Maple}, that the only non-zero component of the Ricci tensor is:
\begin{equation}
R_{uu} = - \left\{ {1\over2} \; g^{AB} \; g_{AB}'' 
- {1\over4} \; g^{AB} \; g_{BC}' \; g^{CD} \; g_{DA}' \right\}.
\end{equation}
Less obviously, a brief computation shows that (up to the usual index symmetries) the only non-zero components of the Riemann tensor are:
\begin{equation}
R_{uAuB} = - \left\{ {1\over2} \; g_{AB}'' 
- {1\over4} \;  \; g_{AC}' \; g^{CD} \; g_{DB}' \right\}.
\end{equation}
This is the same pattern of non-zero components that occurs in the Brinkmann form, which indicates that \emph{some} of our intuition regarding polarization modes will also carry over to the Rosen form of the metric.
Though these formulae look relatively compact, the matrix inversions implicit in ``raising the indices'' mean that these quantities are grossly nonlinear functions of the matrix components $g_{AB}(u)$. In particular, in this form of the metric the $+$ and $\times$ linear polarizations do not decouple in any obvious way.

\subsection{Linear polarization $+$ }

Consider the strong-field gravity wave metric in the $+$ linear polarization. That is, set $g_{xy} = 0$ so that $g_{AB}$ has only two nontrivial components, $g_{xx}$ and $g_{yy}$, corresponding to oscillations along the $x$ and $y$ axes. The resulting metric can be written in the form
\begin{equation}
\label{E:plus}
\d s^2 = - 2\,\d u\;\d v  + f^2(u)  \;\d x^2 + g^2(u) \;\d y^2.
\end{equation}
The only non-zero component of the Ricci tensor is: 
\begin{equation}
R_{uu} = - \left\{ {f''\over f} + {g''\over g}  \right\}.
\end{equation}
Though the expression for the Ricci tensor is compact, ultimately this form of the metric turns out to not be very useful. 
If we write the metric in the form
\begin{equation}
\d s^2 = - 2\,\d u\;\d v + S^2(u) \; \left\{ e^{+X(u)} \;\d x^2 +  e^{-X(u)} \d y^2 \right\},
\end{equation}
then
\begin{equation}
R_{uu} = - {1\over2} \left\{ 4 \; {S''\over S} + {(X')^2}  \right\}.
\end{equation}
Though this may not initially look very promising, it is this version of the metric that permits us to make the most progress. 
In particular note that in vacuum we have
\begin{equation}
X' = 2 \sqrt{-S''/S},
\qquad
\hbox{so that}
\qquad
X(u) = 2 \int^u  \sqrt{-S''/S} \; \d u.
\end{equation}
and the general vacuum wave for $+$ polarization can be put in the form
\begin{eqnarray}
\label{E:3}
\d s^2 &=& - 2\, \d u\;\d v
+ S^2(u) \; \left\{ \exp\left(  2 \int^u  \sqrt{-S''/S} \; \d u \right)  \;\d x^2 \right.
\nonumber
\\
&&
\qquad\qquad\qquad\qquad  \left. +  \exp\left(  -2 \int^u  \sqrt{-S''/S} \; \d u \right) \d y^2 \right\}.
\quad
\end{eqnarray}
Note that as expected from the Brinkmann form we have one free function (per polarization mode).

\subsection{Linear polarization $\times$}

Take the strong-field gravity wave metric in the $\times$ linear polarization
\begin{equation}
\fl \d s^2 = - 2\, \d u\; \d v + {f^2(u)+g^2(u)\over2}\;[ \d x^2 +  \d y^2 ] 
+  [f^2(u)-g^2(u)] \; \d x\; \d y.
\end{equation}
This can be obtained from the $+$ polarization mode (\ref{E:plus}) simply by performing a $45^\circ$ rotation in the $x$--$y$ plane.
The only non-zero component of the Ricci tensor is again: 
\begin{equation}
R_{uu} = - \left\{ {f''\over f} + {g''\over g}  \right\}.
\end{equation}
If we now write this metric in the form
\begin{equation}
\fl \d s^2 = - 2\,\d u\;\d v  + S^2(u) \; \left\{ \cosh(X(u)) \;[\d x^2 + \d y^2] + 2 \sinh(X(u)) \d x \, \d y \right\},
\end{equation}
then, as for the $+$ mode, we have
\begin{equation}
R_{uu} = - {1\over2} \left\{ 4 \; {S''\over S} + {(X')^2}  \right\}.
\end{equation}
In vacuum we can again solve for $X(u)$ and now obtain
\begin{eqnarray}
\d s^2 &=& -2 \,\d u\;\d v  
+ S^2(u) \; \left\{ \cosh\left(  2 \int^u  \sqrt{-S''/S} \; \d u \right) \;[\d x^2 + \d y^2] \right.
\nonumber\\
&&
\qquad \qquad \qquad \left. + 2 \sinh\left(  2 \int^u  \sqrt{-S''/S} \; \d u \right) \d x \, \d y \right\}.
\end{eqnarray}
There is again one freely specifiable function for this $\times$ linear polarization mode. 
By rotating the $x$--$y$ plane through a fixed but arbitrary angle $\Theta_0$ we can easily deal with linear polarization modes along any desired axis, indeed we could have obtained this metric direct from equation~(\ref{E:3}) by a $45^\circ$ rotation in the $x$--$y$ plane. 
The puzzle arises once we try to deal with variable amounts of $+$ and $\times$ polarization simultaneously.

\subsection{Arbitrary polarization}

Let us now take an arbitrary, possibly $u$ dependent polarization, and consider the following metric ansatz:
\begin{eqnarray}
\label{E:gen}
\fl \d s^2 &=& - 2 \, \d u\;\d v  
\nonumber\\
\fl &&
+ S^2(u) \; \left\{ \cosh(B(u)) \;[e^{+X(u)}\d x^2 + e^{-X(u)}\d y^2] + 2 \sinh(B(u)) \d x \, \d y \right\}.
\end{eqnarray}
Note that setting $X(u)=0$ corresponds to $\times$ polarization, while setting $B(u)=0$ corresponds to $+$~polarization. Furthermore we have sufficient free functions, namely \{$S(u)$, $B(u)$, $X(u)$\}, to completely saturate the arbitrary $2\times 2$ symmetric matrix $g_{AB}(u)$.
A brief calculation yields the only nonzero component of the Ricci tensor:
\begin{equation}
R_{uu} = - {1\over2} \left\{ 4 \, {S''\over S} + {(B')^2}  + \cosh^2[B(u)] \;  (X')^2 \right\}.
\end{equation}
Note that $B(u)$ and $X(u)$ have \emph{not} decoupled --- however a hint on how to proceed is provided by noting that the 2-metric
\begin{equation}
\d B^2 +  \cosh^2B  \; \d X^2
\end{equation}
is one of many ways of representing the metric of the hyperbolic plane $H_2$.

\bigskip
\noindent
Now let us try a slightly different representation of the same general metric of equation (\ref{E:gen}):
\begin{eqnarray}
\d s^2 &=& - 2\, \d u\;\d v  + S^2(u) \; \left\{ \vphantom{\Big|} [ \cosh(X(u)) + \cos(\theta(u)) \sinh(X(u))] \d x^2 \right.
\nonumber
\\
&&
\qquad \left. \vphantom{\Big|} + 2 \sin(\theta(u)) \sinh(X(u)) \d x \, \d y   \right.
\nonumber
\\
&&
\qquad \left. \vphantom{\Big|} +  [\cosh(X(u)) - \cos(\theta(u)) \sinh(X(u))] \d y^2 \right\}.
\end{eqnarray}
Note that setting $\theta(u)=0$ corresponds to $+$ polarization, while setting $\theta(u)=\pi/2$ corresponds to $\times$ polarization, and $\theta(u) = \Theta_0/2$ corresponds to linear polarization along axes rotated by an angle $\Theta_0$. Furthermore  we again have sufficient free functions, now \{$S(u)$, $X(u)$, $\theta(u)$\},  to completely saturate the arbitrary $2\times 2$ symmetric matrix $g_{AB}(u)$.
Then
\begin{equation}
R_{uu} = - {1\over2} \left\{ 4 \; {S''\over S} + {(X')^2}  + \sinh^2(X(u)) \;  (\theta')^2 \right\}.
\end{equation}
The vacuum field equations imply
\begin{equation}
  4 \, {S''\over S} + {(X')^2}  + \sinh^2(X(u)) \;  (\theta')^2 = 0.
\end{equation}
Let us introduce a dummy function $L(u)$ and split this into the two equations
\begin{equation}
 4\, {S''\over S} + {(L')^2}   = 0,
\end{equation}
and
\begin{equation}
(L')^2 =  {(X')^2}  + \sinh^2(X(u)) \;  (\theta')^2.
\end{equation}
Ths first of these equations
is just the equation you would have to solve for a pure $+$ or $\times$ or in fact any linear polarization.
The second of these equations can be rewritten as
\begin{equation}
\d L^2 =  \d X^2  + \sinh^2(X) \;  \d\theta^2,
\end{equation}
and is just the statement that $L$ can be interpreted as distance in the 2-dimensional hyperbolic plane $H_2$.

\paragraph{Algorithm:} 
This gives us now a very straightforward algorithm for arbitrary polarization strong-field gravity waves in the Rosen form:
\begin{itemize}
\item Pick an arbitrary $L(u)$ and solve:
\begin{equation}
 4 \; {S''\over S} + {(L')^2}   = 0.
\end{equation}
\item 
Pick an arbitrary curve in the $(X,\theta)$ plane such that $L(u)$ is hyperbolic arc-length along that curve:
\begin{equation}
\d L^2 =  \d X^2  + \sinh^2(X) \;  \d\theta^2.
\end{equation}
\item
This construction then solves the vacuum Einstein equations for the metric
\begin{eqnarray}
\fl \d s^2 &=& - 2\,\d u\;\d v  + S^2(u) \; \left\{ \vphantom{\Big|}  [ \cosh(X(u)) + \cos(\theta(u)) \sinh(X(u))] \d x^2 \right.
\nonumber
\\
\fl &&
\qquad \left.   \vphantom{\Big|}  + 2 \sin(\theta(u)) \sinh(X(u)) \d x \, \d y   \right.
\nonumber
\\
\fl &&
\qquad \left. \vphantom{\Big|}  +  [\cosh(X(u)) - \cos(\theta(u)) \sinh(X(u))] \d y^2 \right\}.
\end{eqnarray}
\end{itemize}
In this sense we have completely solved arbitrary polarization strong-field gravity waves in the Rosen form.

Note the similarities (and differences) with regard to Maxwell electromagnetism, (and with respect to the Brinkmann form).  In Maxwell electromagnetism  the two independent linear polarizations can be specified by 
\begin{equation}
\vec E(u)  = E_x(u) \; \hat x + E_y(u)\; \hat y,
\end{equation}
with no additional constraints (compare with equation (\ref{E:1})). Thus an electromagnetic wavepacket of arbitrary polarization can be viewed as an arbitrary ``walk'' in the $(E_x, E_y)$ plane. We could also go to a magnitude-phase representation $(E,\theta)$ where
\begin{equation}
\vec E(u)  = E(u) \cos\theta(u) \; \hat x + E(u) \sin\theta(u)\; \hat y.
\end{equation}
(Compare with equation~(\ref{E:2}).) 
So an electromagnetic wavepacket of arbitrary polarization can also be viewed as an arbitrary ``walk'' in the $(E, \theta)$ plane, where the $(E, \theta)$ plane is provided with the natural Euclidean metric
\begin{equation}
\d L^2 = \d E^2 + E^2 \; \d \theta^2
\end{equation}
In contrast for gravitational wave in the Rosen form we  are now dealing with an arbitrary ``walk'' in the hyperbolic plane $H_2$, rather than in the Euclidean plane. Furthermore, because of the nonlinearity of general relativity there is still one remaining  differential equation to solve (for the ``envelope''  $S(u)$).

\subsection{Circular polarization}

We can now adopt the above discussion to formulate strong-field circular polarization in the Rosen form. (We emphasize that there is no difficulty whatsoever with weak-field linearized circular polarization, it is only for strong fields that it is difficult to formulate circular polarization for a gravitational wave in the Rosen form.)  Circular polarization corresponds to
\begin{equation}
\theta(u) = \Omega_0\; u; \qquad   X(u) = X_0.
\end{equation}
That is, a fixed distortion $X_0$ with the plane of polarization advancing linearly with retarded time $u$. 
Then
\begin{eqnarray}
\d s^2 &=& - 2\, \d u\;\d v  + S^2(u) \; \left\{  \vphantom{\Big|} [ \cosh(X_0) + \cos(\Omega_0\; u) \sinh(X_0)] \d x^2 \right.
\nonumber
\\
&&
\qquad \left.  \vphantom{\Big|} + 2 \sin(\Omega_0 u) \sinh(X_0) \d x \, \d y   \right.
\nonumber
\\
&&
\qquad \left.  \vphantom{\Big|} +  [\cosh(X_0) - \cos(\Omega_0 u) \sinh(X_0)] \d y^2 \right\}. 
\end{eqnarray}
The only nontrivial component of the Ricci tensor is then
\begin{equation}
R_{uu} = - {1\over2} \left\{ 4 \; {S''\over S}   + \sinh^2(X_0) \;  \Omega_0^2 \right\}.
\end{equation}
The vacuum field equations imply
\begin{equation}
  S''  = -{ \sinh^2(X_0) \;  \Omega_0^2\over 4} \; S,
\end{equation}
whence
\begin{equation}
S(u) = S_0 \; \cos\left\{ { \sinh(X_0) \;  \Omega_0 \; (u-u_0) \over 2}\right\}.
\end{equation}
This now describes a spacetime that has good reason to be called a strong-field circularly polarized gravity wave. Note that the weak-field limit corresponds to $X_0\ll 1$ so that for an arbitrarily long interval in retarded time $u$ we have  $S \approx S_0$, and without loss of generality we can set $S\approx 1$. Then
\begin{equation}
\fl \d s^2 \approx  - 2\, \d u\;\d v  + \d x^2 + \d y^2 + X_0 \; \left\{  \vphantom{\Big|} \cos(\Omega_0\; u) [\d x^2-\d y^2] + 2   \sin(\Omega_0\; u) \;\d x \;\d y \right\}.
\end{equation}
Further generalizations to elliptic polarization are tedious but, given the significantly more general algorithm of the preceding subsection, quite straightforward. 
\subsection{Decoupling the most general Rosen form}

Based on what we have seen so far, one might suspect that there is some general decoupling between the overall ``envelope'' of the gravity wave and the ``directions of oscillation''.  Let us return to considering the metric in general Rosen form
\begin{equation}
\d s^2 = - 2\, \d u\; \d v + g_{AB}(u)\;\d x^A\,\d x^B, 
\end{equation}
where for generality the  $x^A$, $x^B$  represent any arbitrary number of dimensions ($d_\perp\geq 2$) transverse to the $(u,v)$ plane. 
It is easy to check that the only non-zero component of the Ricci tensor is still
\begin{equation}
R_{uu} = - \left\{ {1\over2} \; g^{AB} \; g_{AB}'' 
- {1\over4} \; g^{AB} \; g_{BC}' \; g^{CD} \; g_{DA}' \right\}.
\end{equation}
Let us now decompose the $d_\perp \times d_\perp$ matrix $g_{AB}$ into an ``envelope'' $S(u)$ and a unit determinant related to the ``direction of oscillation''. That is, let us take
\begin{equation}
g_{AB}(u) = S^2(u)\; \hat g_{AB}(u),
\end{equation}
where $\det( \hat g) \equiv 1$. (A related discussion can be found in \S109 of Landau--Lifschitz~\cite{Landau}.) Our goal is to see if we can make the overall ``envelope''  $S(u)$ decouple from $\hat g_{AB}(u)$. 
To start, note that
\begin{equation}
g_{AB}' = 2 S\; S' \; \hat g_{AB} + S^2\; \hat g_{AB}',
\end{equation}
and
\begin{equation}
g_{AB}'' = 2 S\; S'' \; \hat g_{AB} + 2 S'\; S' \; \hat g_{AB}  + 4 S\; S' \; \hat g_{AB}'  + S^2\; \hat g_{AB}''.
\end{equation}
Therefore
\begin{eqnarray}
\fl  g^{AB} \; g_{AB}''  &=& 
  {1\over S^2} \left\{ 2 (S S'' + S' S') d_\perp +  4 S\; S' \; [  \hat g^{AB} \; \hat g_{AB}']  + S^2\; [ \hat g^{AB} \; \hat g_{AB}''] \right\}
\nonumber\\
\fl &=&
 2 \left({ S''\over S}  + {S' S'\over S^2}\right) d_\perp +  4 {S'\over S} \; [  \hat g^{AB} \; \hat g_{AB}']  + [ \hat g^{AB} \; \hat g_{AB}''],
 \end{eqnarray}
 and similarly
 \begin{eqnarray}
\fl  g^{AB} \; g_{BC}' \; g^{CD} \; g_{DA}'  &=&   
  g^{AB} [ 2 S\; S' \; \hat g_{BC} + S^2\; \hat g_{BC}' ] g^{CD} [2 S\; S' \; \hat g_{DA} + S^2\; \hat g_{DA}'] 
 \nonumber\\
\fl &=& 
{1\over S^4} \left\{ 4 (S\; S')^2 d_\perp + 4 S^3 S'  [  \hat g^{AB} \; \hat g_{AB}'] +   S^4 [  \hat g^{AB} \; \hat g_{BC}' \; \hat g^{CD} \; \hat g_{DA}'  ]\right\}
\nonumber\\
\fl  &=& 
 4 \left({S'\over S}\right)^2 d_\perp +  4\left({S'\over S}\right)  [  \hat g^{AB} \; \hat g_{AB}'] +   [  \hat g^{AB} \; \hat g_{BC}' \; \hat g^{CD} \; \hat g_{DA}'  ].
  \end{eqnarray}
Now combine these results
\begin{eqnarray}
\fl R_{uu} &=& 
- \left\{ {1\over2} \; g^{AB} \; g_{AB}'' 
- {1\over4} \; g^{AB} \; g_{BC}' \; g^{CD} \; g_{DA}' \right\}
\nonumber\\
\fl &=& 
- \left({ S''\over S}  + {S' S'\over S^2}\right) d_\perp -2  {S'\over S} \; [  \hat g^{AB} \; \hat g_{AB}']  - {1\over2} [ \hat g^{AB} \; \hat g_{AB}''] 
\nonumber\\
\fl && 
  + \left({S'\over S}\right)^2 d_\perp +  \left({S'\over S}\right)  [  \hat g^{AB} \; \hat g_{AB}'] + 
   {1\over 4}  [  \hat g^{AB} \; \hat g_{BC}' \; \hat g^{CD} \; \hat g_{DA}'  ]
\nonumber\\
\fl &=& 
   - {S''\over S} d_\perp  - {1\over2} [ \hat g^{AB} \; \hat g_{AB}'']  +    {1\over 4}  [  \hat g^{AB} \; \hat g_{BC}' \; \hat g^{CD} \; \hat g_{DA}'  ]
  +  \left({S'\over S}\right)  [  \hat g^{AB} \; \hat g_{AB}']. 
\label{E:111}
\end{eqnarray}
But because we have defined  $\det( \hat g) \equiv 1$ we have as a matrix identity
  \begin{equation}
  [  \hat g^{AB} \; \hat g_{AB}']  = 0,
  \end{equation}
  and differentiating this one more time
  \begin{equation}
   [  \hat g^{AB} \; \hat g_{AB}''] -  [  \hat g^{AB} \; \hat g_{BC}' \; \hat g^{CD} \; \hat g_{DA}'  ] = 0.
  \end{equation}
Therefore 
\begin{equation}
R_{uu} =  - d_\perp \,  {S''\over S}   - {1\over2}\, [  \hat g^{AB} \; \hat g_{BC}' \; \hat g^{CD} \; \hat g_{DA}'  ],
\end{equation}
or more abstractly
\begin{equation}
R_{uu} =  - d_\perp  \, {S''\over S}  - {1\over2} \; \tr\left\{  [\hat g]^{-1} \; [\hat g]' \; [\hat g]^{-1} \; [\hat g]'  \right\}. 
\end{equation}
Note that we have now succeeded in decoupling  the determinant ($\det(g) = S^{2d_\perp}$; effectively the  ``envelope'' $S(u)$ of the gravitational wave) from the unit-determinant matrix $\hat g(u)$. This observation is compatible with all the specific examples considered above.
Now consider the set $SS(I\!\!R, d_\perp)$ of all unit determinant real symmetric matrices, and on that set (not a group) consider the Riemannian metric
 \begin{equation}
 \d L^2 =  \tr\left\{  [\hat g]^{-1} \;\d[\hat g] \; [\hat g]^{-1} \; \d[\hat g]  \right\}.
\end{equation}
Then
\begin{equation}
R_{uu} =  -  {1\over2} \left\{ 2 d_\perp\, {S''(u)\over S(u)}   + \left({\d L\over\d u}\right)^2  \right\}. 
\end{equation}
The vacuum Einstein equations then reduce to
\begin{equation}
{\d L\over\d u} = \sqrt{ - 2 \d_\perp \; {S''\over S}};   \qquad  L(u) = \int^u  \sqrt{ - 2 \d_\perp \; {S''\over S}} \; \d u.
\end{equation}
That is, an arbitrary polarization vacuum Rosen wave is a random walk in $SS(I\!\!R, d_\perp)$,  with distance along the walk $L(u)$ being related to the envelope function  $S(u)$ as in the discussion above. 

To probe the polarization modes in more detail, a completely analogous but slightly more complicated calculation determines the Riemann tensor components to be
\begin{eqnarray}
\fl
R_{uAuB} = - \left\{
  S S'' \hat g_{AB} +S^2 \left[  {1\over2} \; \hat g''_{AB}  -{1\over4}   \; \hat g_{AC}' \; \hat g^{CD} \; \hat g_{DB}'  \right] +  S S' \hat g'_{AB}
  \right\}.
\end{eqnarray}
Note that this is compatible with equation (\ref{E:111}).

\section{Discussion}

Strong field gravitational waves, both in Brinkmann~\cite{Brinkmann} form and Rosen~\cite{Rosen} form, have been known for some 85 years. Despite this there are still a number of surprises hiding in the Rosen form.  In particular, while arbitrary time-dependent polarization mixtures are trivial to deal with in the Brinkmann form they appear to be much more difficult to implement in the Rosen form.  To address this  puzzle we have re-analyzed the Rosen gravity wave in terms of an ``envelope'' function and two freely specifiable functions corresponding to the individual polarization modes. The vacuum field equations can be reinterpreted in terms of a single differential equation governing the ``envelope'' coupled with what is essentially a random walk in polarization space.  In particular we have indicated how to construct a circularly polarized strong field Rosen form gravity wave,  and how to generalize this central idea beyond (3+1) dimensions.

\section*{Acknowledgements}

This research was supported by the Marsden Fund administered by the Royal Society of New Zealand.

\section*{References}



\end{document}